\documentclass[]{rsos}%%%%where rsos is the template name

\usepackage[numbers]{natbib}

\begin{document}

%%%% Article title to be placed here
\title{A data-driven model for influenza transmission incorporating media effects}

\author{%%%% Author details
Lewis Mitchell \& Joshua V. Ross}

%%%%%%%%% Insert author address here
\address{%                             % unique id
  School of Mathematical Sciences,
  University of Adelaide, % university, etc
  North Terrace,                       %
  5005                               % post or zip code
  Adelaide,                              % city
  Australia                               % country
}

%%%% Subject entries to be placed here %%%%
\subject{applied mathematics, health and disease and epidemiology, mathematical modelling}

%%%% Keyword entries to be placed here %%%%

%%%% Insert corresponding author and its email address}
\corres{Lewis Mitchell\\
\email{lewis.mitchell@adelaide.edu.au}}

%%%% Abstract text to be placed here %%%%%%%%%%%%
\begin{abstract}
Numerous studies have attempted to model the effect of mass media on the transmission of diseases such as influenza, however quantitative data on media engagement has until recently been difficult to obtain.  With the recent explosion of ``big data'' coming from online social media and the like, large volumes of data on a population's engagement with mass media during an epidemic are becoming available to researchers. In this study we combine an online data set comprising millions of shared messages relating to influenza with traditional surveillance data on flu activity to suggest a functional form for the relationship between the two.
Using this data we present a simple deterministic model for influenza dynamics incorporating media effects, and show that such a model helps explain the dynamics of historical influenza outbreaks. Furthermore, through model selection we show that the proposed media function fits historical data better than other media functions proposed in earlier studies.
\end{abstract}
%%%%%%%%%%%%%%%%%%%%%%%%%%%

%%%%%%%%%% Insert the texts which can accomdate on firstpage in the tag "fmtext" %%%%%

\begin{fmtext}

\section{Introduction}

Traditional models of epidemics assume static parameter values over the course of an outbreak \cite{KR2007}. 
As such, they do not allow for changes in human behaviour which in turn are likely to impact the rate of transmission in a population.  Such behavioural changes in response to disease outbreaks are well established 
\cite{Funk2010}.  This includes self-imposed social distancing during influenza pandemics \cite{Epstein2008}, 
and the usage of face masks and changes in travel behaviour during the Severe Acute Respiratory Syndrome (SARS) outbreak of 2002--2004 \cite{Bell2004}.  The term {\em prevalence elastic behaviour} has arisen to explain voluntary protective behaviour which increases with disease prevalence \cite{Geoffard1996}, as has been observed for both measles \cite{Philipson1996} and HIV \cite{Kremer1996}.

\end{fmtext}

\maketitle

%%%%%%%%%%%%%%% End of first page %%%%%%%%%%%%%%%%%%%%%

The close to real-time awareness of disease prevalence in an outbreak is now common due to the relatively recent explosion in mass and social media. 
The past decade has seen significant growth in studies concerning the interaction of media, 
human behaviour 
and infectious disease dynamics, 
and there now exists a substantial body of work on this topic \cite{Ginsberg2009,Culotta2010,Funk2010,Lampos2010,Signorini2011,Lamb2013,Collinson2015,Zhang2015}. 
Despite this growth,
empirical studies of prevalence elastic behaviour due to mass media have until recently been difficult due to the lack of availability of data directly 
measuring media engagement
and relating it to behavioural change.
As such, 
the vast majority of studies in this area can be broadly classified into two groups,
with slightly different motivations. 
First,
{\em pure mathematical models of behavioural change}, 
in which a model is formulated that accounts for how dynamics are influenced by disease awareness or prevalence, 
typically facilitated through media 
-- 
these are often either in the form of introducing new states which account for the behavioural status of individuals \cite{DOnofrio2007}, 
by allowing modification to the contact structure \cite{Epstein2008,Greenhalgh2015}, 
or by allowing modification to the model parameters \cite{Kiss2010,Xiao2015}
-- 
and the consequences are then explored.
Collinson et al. (2015) model behavioural change due to media by explictly including a compartment for individuals influenced by mass media into an SEIR-type model,
also incorporating effects like vaccination and social distancing \cite{Collinson2015}.
This study is of particular interest due to the fact that it incorporates a ``media fatigue'' effect during the 2009/10 H1N1 pandemic by fitting to news report data collected from newspaper homepages during the pandemic. 
Second, 
{\em pure statistical models of media and prevalence}
 are used on large data sets to produce statistical regression models relating some measure of volume of media concerning epidemics to the prevalence of infection \cite{Culotta2010,Lampos2010} or reproductive number \cite{Majumdar2015}.
Such models have recently become popular due to the rapid increase in new data streams coming from internet and online social media usage \cite{Achrekar2011,Broniatowski2013}.
The study of Signorini et al. (2011) is an exception to this trend: 
whilst it is a pure statistical model, 
it includes an investigation of the relationship between "tweets" on Twitter and public sentiment with respect to H1N1 \cite{Signorini2011}.
The FluOutlook platform \cite{Zhang2015} is also particularly interesting;
by using a variety of data sources, including Twitter, to initialise a global agent-based epidemiological model it is able to produce real-time forecasts of an evolving influenza season.

Here our focus is on simple models for incorporating behavioural changes from awareness of disease prevalence, 
through modification to the effective transmission rate parameter. 
We measure disease dynamics through influenza incidence data from the United States over the period 1998/99--2014/15,
and human behaviour through social media data collected from Twitter over the period 2009/10--2014/15. 
Modification to the effective transmission rate is via a so-called 
{\em media function}. 
Three distinct media functions have been introduced, 
and recently compared, 
in the literature \cite{Collinson2014}. 
A potential criticism of pure mathematical model based studies, 
as described above, 
is that the usefulness of the model when analysing real data is uncertain. 
In fact, 
as we will show here, 
some of these models have only very limited use in describing data coming from historical influenza outbreaks. 
On the other hand, 
whilst pure statistical models of media and prevalence are potentially of use for detecting and tracking disease incidence, 
they are subject to typical criticisms of ``big data'' analyses \cite{Lazer2014}
as containing biases and tending towards overfitting.
As such,
their usefulness for understanding potential mechanisms of impacts, 
as is the focus of model-based analyses, 
is limited. 

We propose a data-driven approach that couples these existing paradigms:
through a statistical analysis of data on media engagement and disease prevalence
we develop a mathematical model of behaviour change which may then be validated against data.
Our approach uses online social media data from Twitter 
alongside surveillance data on influenza 
to inform the form of the media function. 
The motivation is that by using both sources of data we 
have some empirical justification for the form of the chosen media function
and can also better describe real observations. 
By using model selection criteria, 
we show that the media function proposed here fits historical surveillance data better than other media functions proposed in earlier studies.

The structure of the remainder of this paper is as follows:
in Section \ref{sec:methods} we describe the data set and model used,
in Section \ref{sec:results} we show results comparing our proposed model with 
surveillance data,
and then conclude with a discussion in Section \ref{sec:discussion}.

\section{Methods}
\label{sec:methods}

In order to measure media engagement we use a corpus 
of over 2.9 million geolocated, flu-related tweets collected from the contiguous United States between September 2009 and July 2015.
This sample was provided by the Computational Story Lab at the University of Vermont,
and is a subset of Twitter's ``garden hose'' feed,
representing roughly 10\% of all public messages posted to the platform.
In the present study we consider only tweets which contain one or more of the strings `flu',
`\#flu',
`influenza'
or `\#influenza'.
Furthermore,
we will focus on ``retweeted'' messages,
where an individual has opted to reshare a tweet originally authored by someone else with their own followers by means of a retweet button within the Twitter interface or by appending the string `RT' to the beginning of the original message.
Such messages account for approximately 30\% of the corpus
and are mainly resharings of flu-related articles from major news outlets,
but can also contain retweets of messages authored by regular Twitter users.

We use a deterministic SEEIIR-M model 
({\it susceptible--exposed--infected--recovered with media},
with two compartments for exposed and infected individuals)
to model the transmission of influenza under the influence of media effects:
\begin{align}
\label{eqn:S}\dot{S}   &= -\beta f(I) SI\\
\dot{E}_1 &= \beta f(I)SI - 2\sigma E_1\\
\dot{E}_2 &= 2\sigma E_1 - 2\sigma E_2\\
\dot{I}_1 &= 2\sigma E_2 - 2\gamma I_1\\
\dot{I}_2 &= 2\gamma I_1 - 2\gamma I_2\\
\label{eqn:R}\dot{R}   &= 2\gamma I_2
\end{align}
where 
$S,E_1,E_2,I_1,I_2$ and $R$ represent the proportions of the population in each compartment,
$S + E_1 + E_2 + I_1 + I_2 + R = 1$,
$\beta$ represents the effective transmission rate in the absence of media effects,
$1/\sigma$ represents the average latent period,
$1/\gamma$ represents the average infectious period,
and $f(I)$ is the so-called {\it media function}
which represents the reduction in transmission of the disease through the influence of mass media.
Consequently, $0 \leq f(I) \leq 1$ 
with $f(I) \equiv 1$ implying no effect of media upon transmission,
and we will assume $f(I)$ is monotonically decreasing in $I$.
Setting $f(I) \equiv 1$ recovers the standard SEEIIR model.
As $f(0) = 1$ for each media function, 
the basic reproduction number 
$R_0 = \frac{\beta}{\gamma}$, 
which is independent of $f(I)$.
The two compartments for the exposed and infectious periods mean that these periods have underlying Erlang-2 distributions with mean exposed and infectious periods $1/\sigma$ and $1/\gamma$ respectively,
which have been shown to more accurately represent the shape of observed distributions \cite{Wearing2005}.
We found similar results using standard SEIR-type models;
these results are presented in the Appendix.
Note that we have not included vaccination in our model, for two reasons:
firstly, for comparison with media models from previous studies (see below) which use SEIR-type models without vaccination;
and secondly, because vaccination coverage in adults has remained approximately constant since 2010 \cite{CDC} and the Twitter data we will study primarily relates to media reporting around the peak of the influenza season rather than the earlier peak of the vaccination season.
Our model therefore can essentially be considered as a model for influenza dynamics in the unvaccinated portion of the population.

Previous studies have postulated a number of different forms for $f(I)$;
see \cite{Collinson2014} for a recent review.
In particular,
\cite{Cui2007} set
\begin{equation}
\label{eqn:f1}f_1(I) = \exp(-p_1 I)
\end{equation}
within an SEI model,
\cite{Xiao2007} used 
\begin{equation}
f_2(I) = \frac{1}{1 + p_2 I^2}
\end{equation}
within an SIR model to account for the psychological effects of a large population infected with SARS,
and many authors (for example \cite{Cui2008,Tchuenche2011}) set
\begin{equation}
\label{eqn:f3}f_3(I) = \frac{1}{1 + p_3 I}
\end{equation}
to account for various effects including media coverage.

To compare the model outputs with real data we use
influenza surveillance data provided by the US Centre for Disease Control (CDC) \cite{CDC}.
Specifically,
we fit models to the nation-wide percentage of new laboratory-confirmed influenza cases per week.
We find best fits for the free model parameters to the surveillance data by minimising least-square error between model solutions and surveillance data using a 
limited memory Broyden-Fletcher-Goldfarb-Shanno (L-BFGS-M) method \cite{Zhu1997},
implemented in Python.
To ensure the numerical stability of the numerical optimisation routine, 
we constrain $R_0$ to be between 1 and 2, 
the mean infectious period $1/\gamma$ to be between 1 and 5 days, 
and the mean exposed time $1/\sigma$ to be between 1 and 3 days.
To perform model selection we use the Akaike Information Criterion (AIC) with finite sample size correction.

% \note{will minimise least-square error wrt CDC data to find best model 
% (L-BFGS-M? method),
% select models based on AIC with finite sample size correction.}

\section{Results}
\label{sec:results}

% \note{data suggests that media engagement may be linear wrt proportion of population infected}

\begin{figure}[h!]
  \includegraphics[width=\columnwidth]{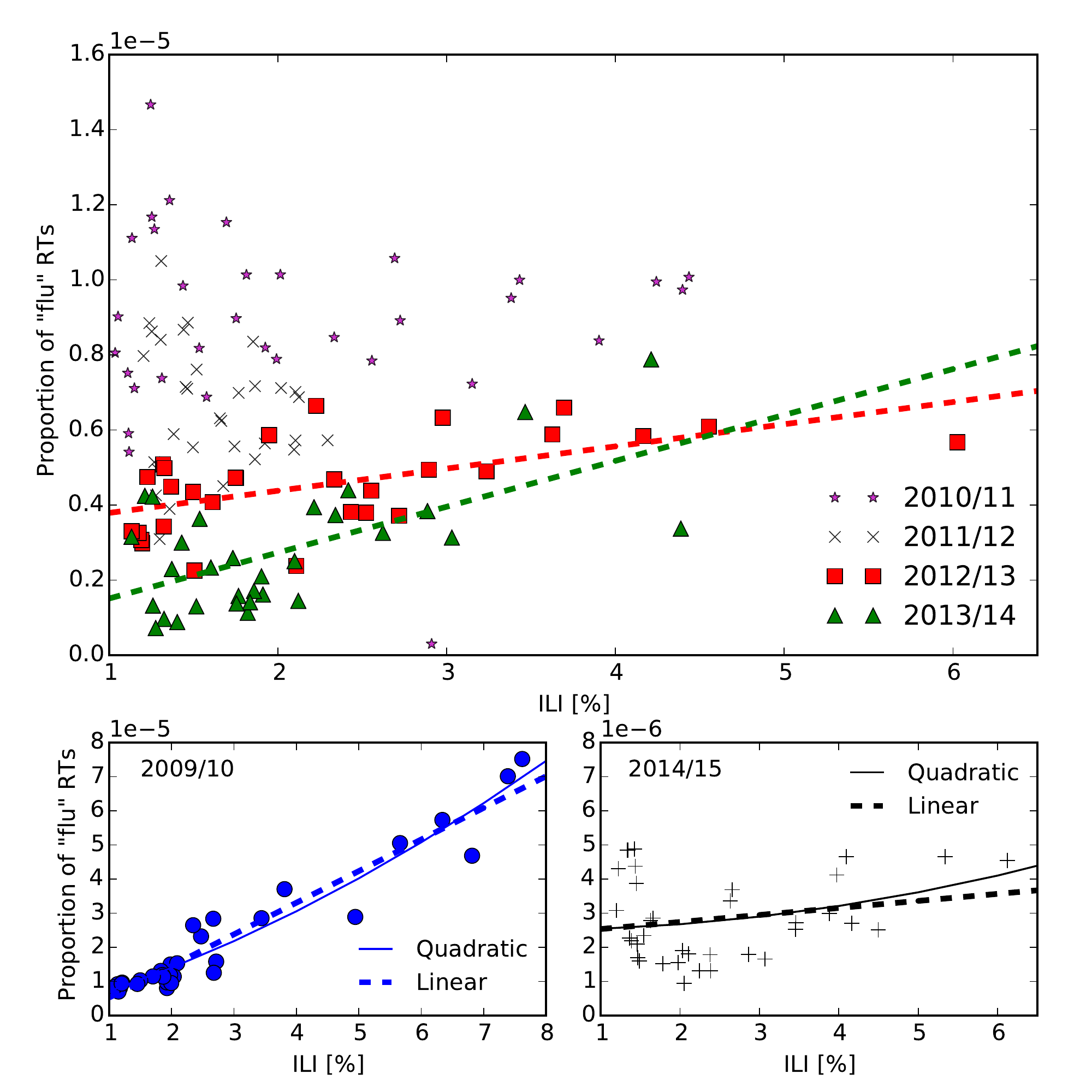}
  \caption{Media engagement from Twitter data.
      Correlation between proportion of public retweets regarding 'flu' and number of influenza-like-illness (ILI) cases, 2009/10-2014/15.
      ILI data is expressed as a percentage of the total number of visits to sentinel surveillance providers.
      Linear trend lines are shown for the years showing significant ($p < 0.01$) correlation.
      The subfigures show both quadratic and linear fits to the data for the 2009/10 and 2014/15 seasons.}
      \label{ILI_RT_USA}
      \end{figure}

We use the act of sharing a message pertaining to influenza as a proxy for an individual engaging with media about an influenza outbreak.
While this act of sharing does not necessarily imply that the individual will change their behaviour,
it does suggest that the individual is at least somewhat {\it concerned} by the media surrounding the influenza outbreak.
Figure \ref{ILI_RT_USA} shows the relationship between proportion of US-based tweets which were retweets concerning influenza
(that is,
number of retweets containing one or more of the strings
`influenza',
`flu'',
`\#influenza', or
`\#flu'
divided by the total number of tweets)
and the number of ILI cases per week
for the 2009/10 to 2014/15 influenza seasons,
expressed as a percentage of the total number of visits to sentinel providers.
The data on weekly counts of ILI activity and retweeting rates used can be found in the Electronic Supplementary Material.
We chose to fit to ILI activity rather than laboratory-confirmed influenza incidence because we expect individuals to tend to share flu-related information on social media upon feeling ill,
rather than strictly once they are confirmed to have influenza.
The 2009/10 pandemic (plotted in the lower left subplot) stands out as having the largest number of both ILI cases and retweet activity.
We observe strong Pearson correlations between retweets and influenza activity for 3 out of the 6 years plotted 
--
in 2009/10, 2012/13 and 2013/14 ($p < 0.01$).

% \note{hence, hypothesis: $f(I) = 1 - p_m I$.}
Importantly,
while the relationship between media engagement and flu activity is small,
it is roughly linear for most flu seasons plotted.
Using AIC to test linear and quadratic models for the data,
we found that the linear model was selected in all seasons apart from 2009/10.
We show linear and quadratic fits for this season as well as 2014/15
in the subplots below the main figure.
In 2014/15 the linear model was slightly preferred with a relative Akaike weight of 0.58 to 0.42 for the quadratic model,
while in 2009/10 the quadratic model was slightly preferred with a relative Akaike weight of 0.55 to 0.45 for the linear model.
Note that as demonstrated by the model fits in the two subfigures,
the Akaike weights indicate that there is substantial support in the data for both the linear and quadratic models.
Indeed, 
we found that the relative likelihood of the quadratic model increased with the total number of ILI cases per season (see Appendix),
suggesting that nonlinear media effects may become increasingly relevant during more severe outbreaks.
We also present residual plots for the linear and quadratic models for all years in the Appendix, 
showing no obvious non-random patterns for the model fits,
along with further details of the AIC model selection and a table of relative Akaike weights for all years.
Note also that we observed similar-looking relationships between media engagement and influenza activity when using the number of comments on flu-related articles in the New York Times between 2001 and 2013 as our metric for media engagement.
However, due to the smaller amount of data we could only find a statistically significant correlation between the two during the 2009 pandemic.

Based upon these observations and for simplicity in comparing models,
we propose the following simple linear media function to describe the reduction in transmission due to media effects:
\begin{equation}
\label{eqn:fm}f_m(I) = 1 - p_m I
\end{equation}
where $p_m$ is a parameter (to be fitted) describing the reduction in actual transmission transmission due to concern from media coverage. 
Yorke and London applied a similar function in a different context,
to model exposure rates for seasonal measles outbreaks \cite{Yorke1973}.
% \note{(need to phrase this carefully/precisely.)}
Note that in order to assure that
$ 0 \leq f_m(I) \leq 1$
it will be necessary to constrain $p_m$ such that 
$ 0 \leq p_m \leq 1$, as $ I \leq 1$ always.
This is in contrast to the media functions (\ref{eqn:f1})--(\ref{eqn:f3}),
for which the parameters can take on any value $p_1,p_2,p_3 \geq 0$.
We remark that while an obvious extension for larger outbreaks would be to use a quadratic media function
$f_m(I)  = 1 - p_{m1}I - p_{m2}I^2$,
for ease of comparison with existing media functions we will only consider the one-parameter model (\ref{eqn:fm}).

% \note{present some sample time series showing how the effect of this media function
% on the outbreak dynamics}

\begin{figure}[h!]
    \includegraphics{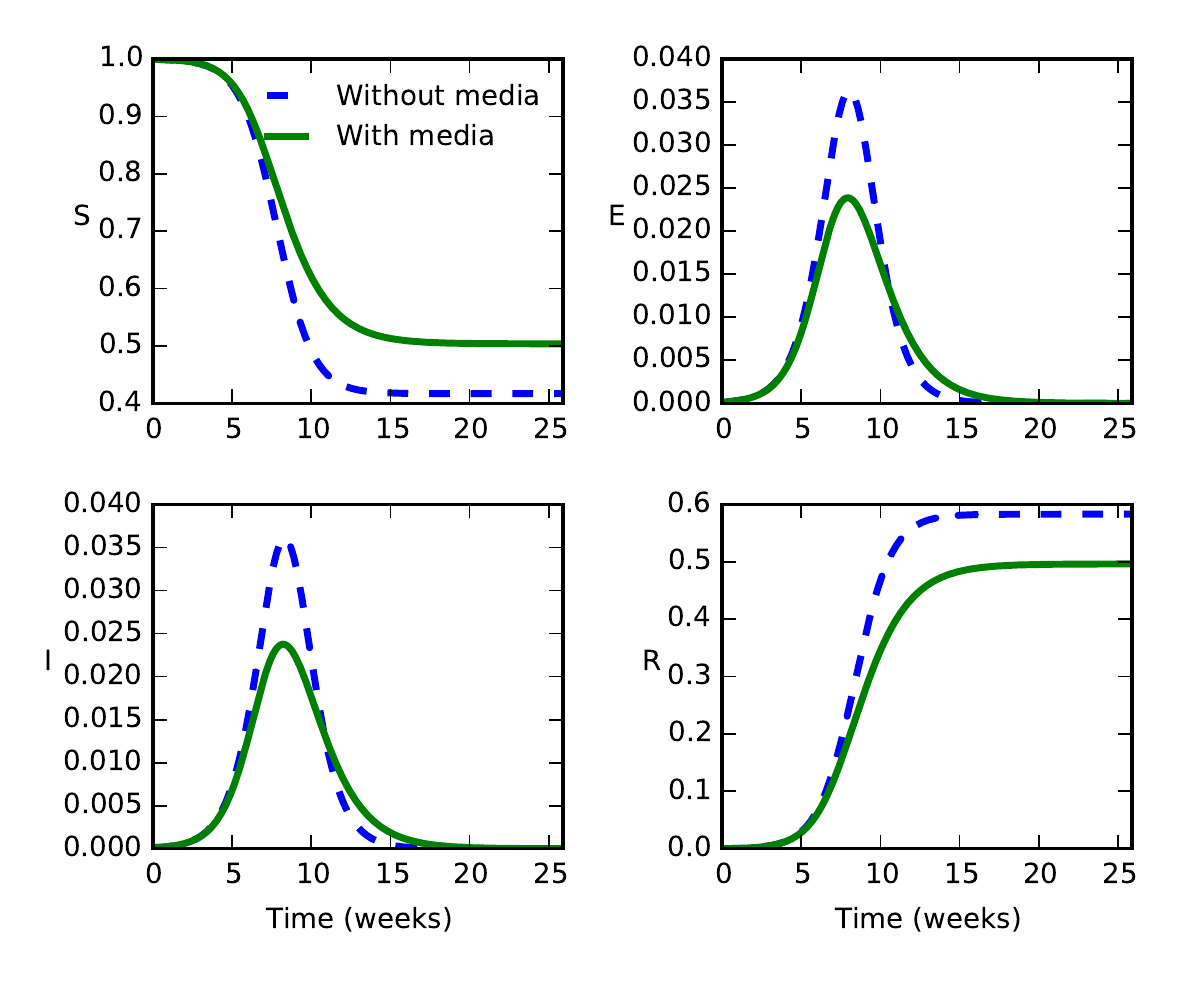}
\caption{Sample time series showing the effect of the media function $f_m(I) = 1 - p_m I$.
The media function reduces the final epidemic size,
and slows the decay rate of the outbreak.}
\label{fig:samplePlots}
  \end{figure}

We show an example of the effect of the media function $f_m$ upon the dynamics in Figure \ref{fig:samplePlots},
where we have set 
$p_m = 0.05$,
$R_0 = 1.5 $,
$\gamma = 1/2$ (days)$^{-1}$,
$\sigma = 1/2$ (days)$^{-1}$
and have plotted $E = E_1 + E_2$ and $I = I_1 + I_2$.
The media function reduces the total number of infected persons (i.e., the final size of the epidemic)
and size of the peak,
while not noticeably changing the timing of the peak.
The slower rate of depletion of susceptibles means that the infection dies out slightly slower in the model with media effect.

% \note{we minimise least squares error wrt data in order to get the following model parameters}

To investigate how well the various transmission models, both with and without media effects, describe real influenza outbreaks,
we fit (\ref{eqn:S})-(\ref{eqn:R}) with $f(I)\equiv1$ as well as (\ref{eqn:f1})-(\ref{eqn:fm}) to weekly laboratory-confirmed influenza incidence data for the 1998--2013 flu seasons using least squares.
Note that unlike social media engagement which can be reasonably expected to relate to ILI,
it is appropriate to fit models of the underlying disease dynamics to confirmed influenza incidence data only.
Using the L-BFGS-B method, 
we find parameter values 
$R_0$, 
$\sigma$, 
$\gamma$, 
and media parameter $p_m$
which best fit the data.
The best-fitting parameters for each model for the 2013/14 flu season are shown in Table \ref{tbl:params},
and for all other seasons are shown in the Appendix.
We fit observations from 4 weeks before the peak to 12 weeks after the peak.
Also shown in Table \ref{tbl:params} are the average conditional probabilities for each model,
as obtained from the normalised Akaike weights for each model across all flu seasons between 1998/99 and 2014/15 in which a non-zero media function was found.

\begin{table}[h!]
\caption{Parameters of best fit for SEEIIR and SEEIIR-M models for 2013/14 influenza season.
% \note{NOTE: is this dicey? the media coverage has effectively reduced $R_0$. 
% this makes sense within the model, but is it biologically reasonable?
% also, the latent period and recovery rates seem a bit fast. 
% perhaps these should be fixed? 
% they should be physcial parameters which don't depend on any media effects.
% then again -- presumably they change from season to season,
% and there's no reason to think that the values estimated from one model are any more correct than from the other.}
\label{tbl:params}
}
      \begin{tabular}{cccccc}
        \hline
           & $f(I) \equiv 1$  & $f_m$ & $f_1$ & $f_2$ & $f_3$\\ 
           \hline
        $R_0$    & 1.1574 & 1.5101  & 1.8574 & 1.4949 & 1.9281 \\
        $1/\sigma$ (days) & 1 & 1.6881 & 2.2162 & 1.9873 & 1.3794\\
        $1/\gamma$ (days) & 1.1979 & 1 & 1.2719 & 1.0979 & 1.1001 \\
        $p_i$ & --  & 0.3316 & 0.1543 & 0.7381 & 0.8140\\ 
        \hline
        $p_{AIC}$ & $O\left(10^{-9}\right)$ & $>0.9999$ &$O\left(10^{-8}\right)$ & $O\left(10^{-7}\right)$& $O\left(10^{-5}\right)$\\
        \hline
      \end{tabular}
      
\end{table}

\begin{figure}[h!]
  \includegraphics[width=\columnwidth]{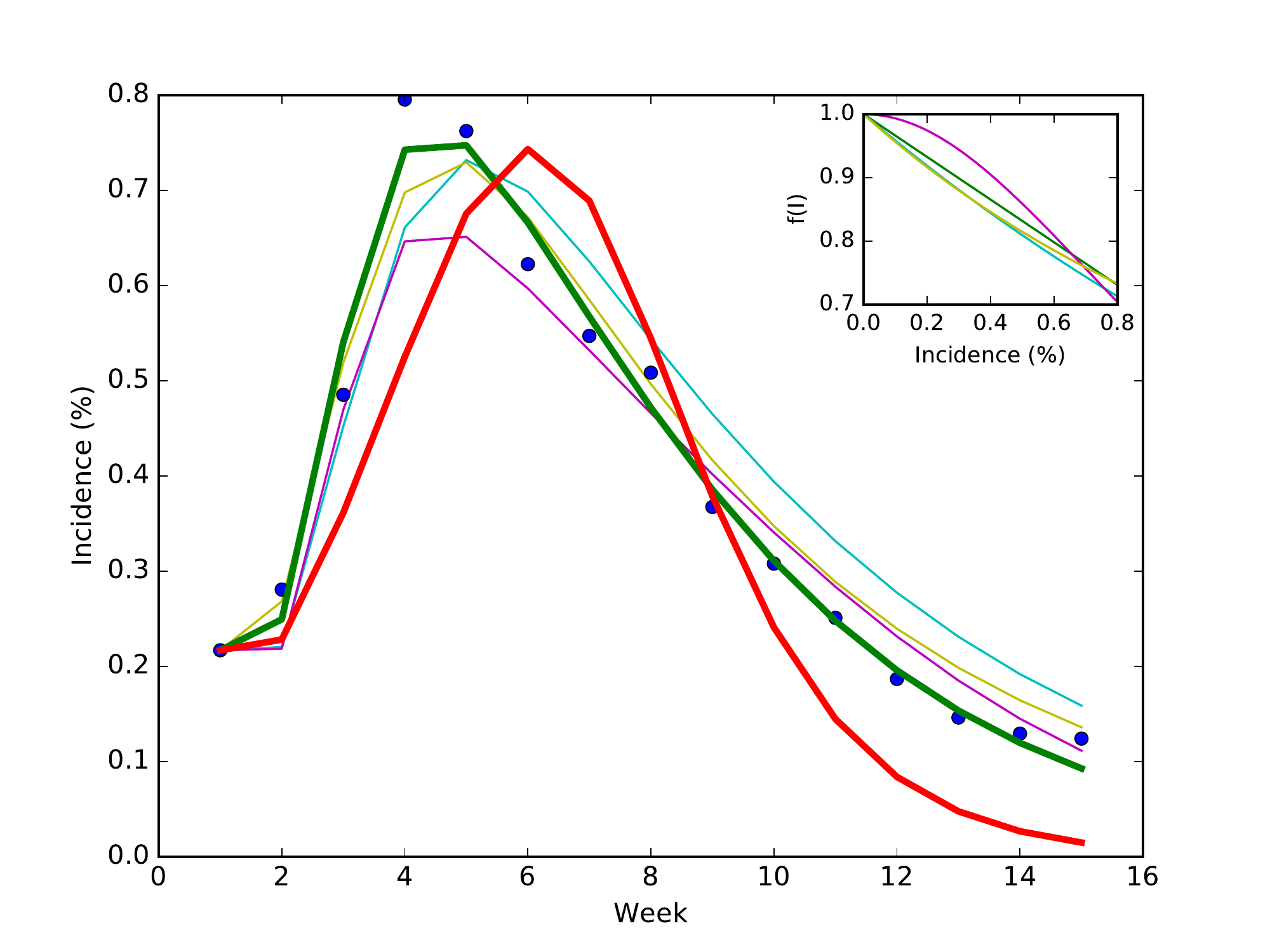}
  \caption{Best model fits to data from 2013/14 flu season.
      Weekly laboratory-confirmed influenza incidence data (blue dots), and model fits for SEEIIR model without media function (red), with media function (green), and variations 
$f_1$ (cyan),
$f_2$ (magenta)
and $f_3$ (yellow),
for the 2013/14 influenza season (USA).
Incidence data is normalised by the total number of visits to sentinel providers.}
      \label{fig:2013}
      \end{figure}

In Figure \ref{fig:2013} we show example fits to observations of the percentage of new laboratory-confirmed influenza cases per week (blue) for the model with 
no media effect (red)
and media functions given by
$f_m$ (green),
$f_1$ (cyan),
$f_2$ (magenta)
and $f_3$ (yellow)
for the 2013/14 influenza season.
As with the ILI data, the laboratory-confirmed case data is expressed as a percentage of the total number of visits to sentinel surveillance providers.
The inset plot shows the corresponding media functions with best-fitting parameters.
While no model is able to correctly estimate the size of the peak of the infection,
the model with linear media function $f_m$ is the only one which correctly identifies the week in which the infection peaks.
The media functions $f_m$ and $f_1$ are also the only models which describe the decay of the infection post-peak well.

% \note{remark about fitting the post-peak period}

While the media function $f_m(I)$ was derived based upon Twitter data, 
our intention in focussing on news-sharing behaviours is to model the effects of mass media more generally.
Indeed, we might expect that population-level engagement with other forms of mass media show a similar monotonically decreasing relationship between media coverage and transmission.
To that end we now apply the proposed media function to all influenza seasons we have incidence data for, 
and find similar results for most seasons between 1998/99 and 2014/15. 
Table \ref{tbl:probs} shows the average conditional probability of selecting each model,
where the average is taken over all years in which a media function is required at all.
Also shown are the 95\% confidence intervals for each average conditional probability. 
No media functions of any kind were required to describe the 2003/04 flu season,
$f_2$ gave the best fit to observations in 2006/07 only, 
and $f_1$ gave the best fit in 2009/10 only.

\begin{table}[h!]
\caption{Average probability of selecting each model over the 1998/99-2014/15 seasons.
We have fit over a 16-week period for each season.
The 2009/10 pandemic influenza season has been excluded.}
\label{tbl:probs}
      \begin{tabular}{lcc}
        \hline
           &  $p_{AIC}$ & 95\% CI  \\ 
           \hline
        $f(I)\equiv 1$                       & 0.0500 & [0,0.1398] \\
        $f_m(I) = 1 - p_m I$            & 0.8347 & [0.6674,1] \\
        $f_1(I) = \exp(-\gamma p_1 I)$   & 0.0267 & [0,0.0535] \\
        $f_2(I) = \frac{1}{1 + p_2 I^2}$ & 0.0060 & [0.0001,0.0120] \\ 
        $f_3(I) = \frac{1}{1 + p_3 I}$   & 0.0826 & [0,0.1871]\\
        \hline
      \end{tabular}
\end{table}

\begin{figure}[h!]
    \includegraphics{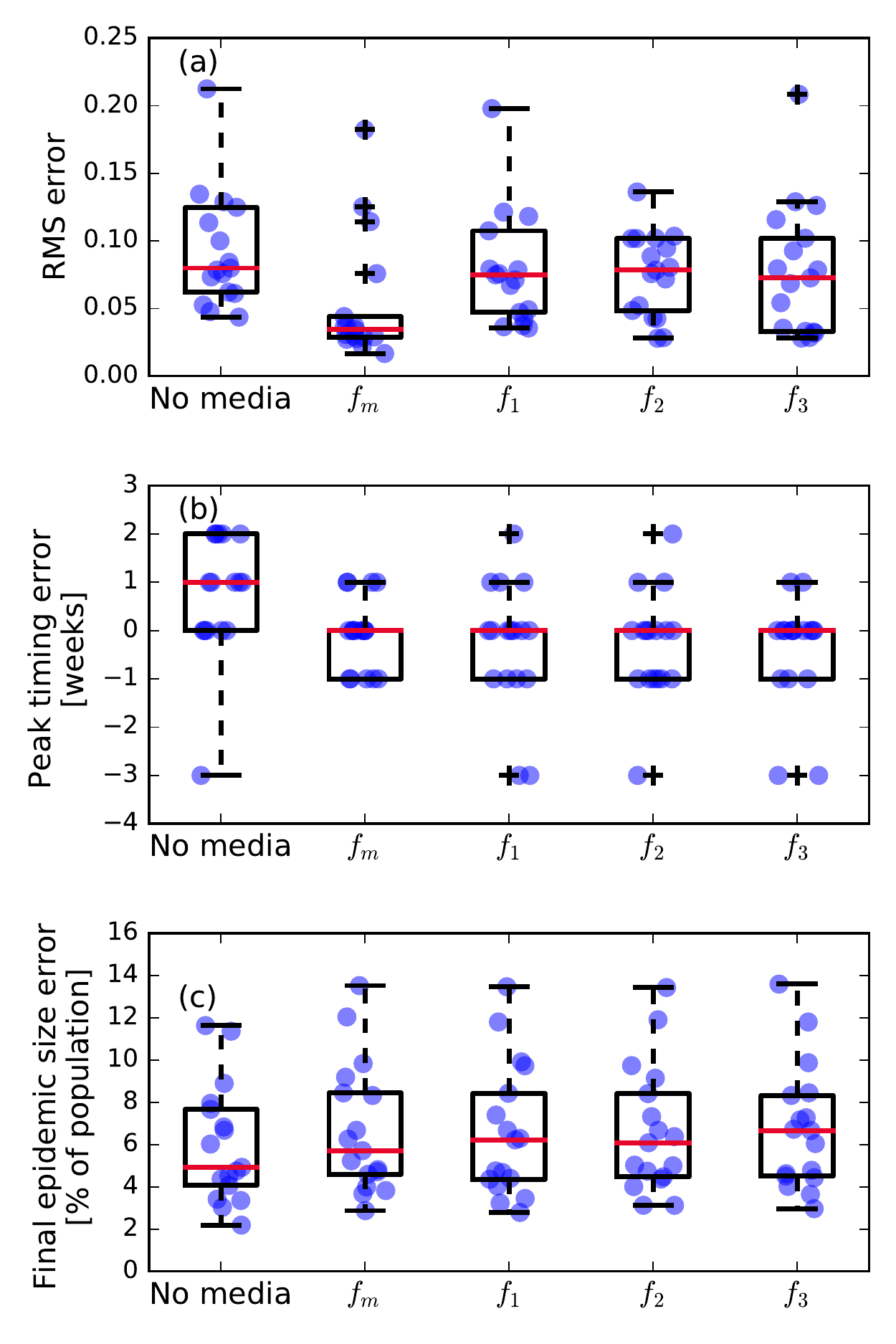}
    \caption{Boxplots of RMS error (a), 
    peak timing error (b), 
    and final epidemic size error (c),
    for standard model and models with media effects for 1998/99-2014/15 seasons.
    The 2009/10 pandemic influenza season has been excluded.
    Blue dots show results from individual years (where we have added random jitter for visibility),
    crosses show outliers.
    }
    \label{fig:boxplots}
\end{figure}

We next examine how well models with and without media function estimate
the complete epidemic curve,
as well as the peak timing and severity.
Figure \ref{fig:boxplots} shows boxplots of 
(a) RMS error,
(b) peak timing error,
and (c) final epidemic size error,
for the model with no media effect
as well as media functions as defined in
(\ref{eqn:f1})-(\ref{eqn:fm})
over the 1998/99--2014/15 seasons.
The proposed media function $f_m$ significantly outperforms all other models
with or without media effects at fitting the epidemic curve,
with the distribution of RMS errors significantly less spread and centred closer to 0 than all other models.
All models with media effect are significantly better than the standard model at matching the observed peak timing of an outbreak (Mood's median test, $p=0.05$),
although there is no significant difference between the four models.
Similarly,
there is no significant difference across models in explaining the observed final epidemic size,
in fact the median error for the standard model without media effect is slightly lower than that for the models with media functions (however, this difference is not significant).

% \note{media function is most useful at explaining the period post-peak.
% quality of model fit decreases as number of weeks leading up to peak increases.}

\begin{figure}[h!]
  \includegraphics[width=0.8\columnwidth]{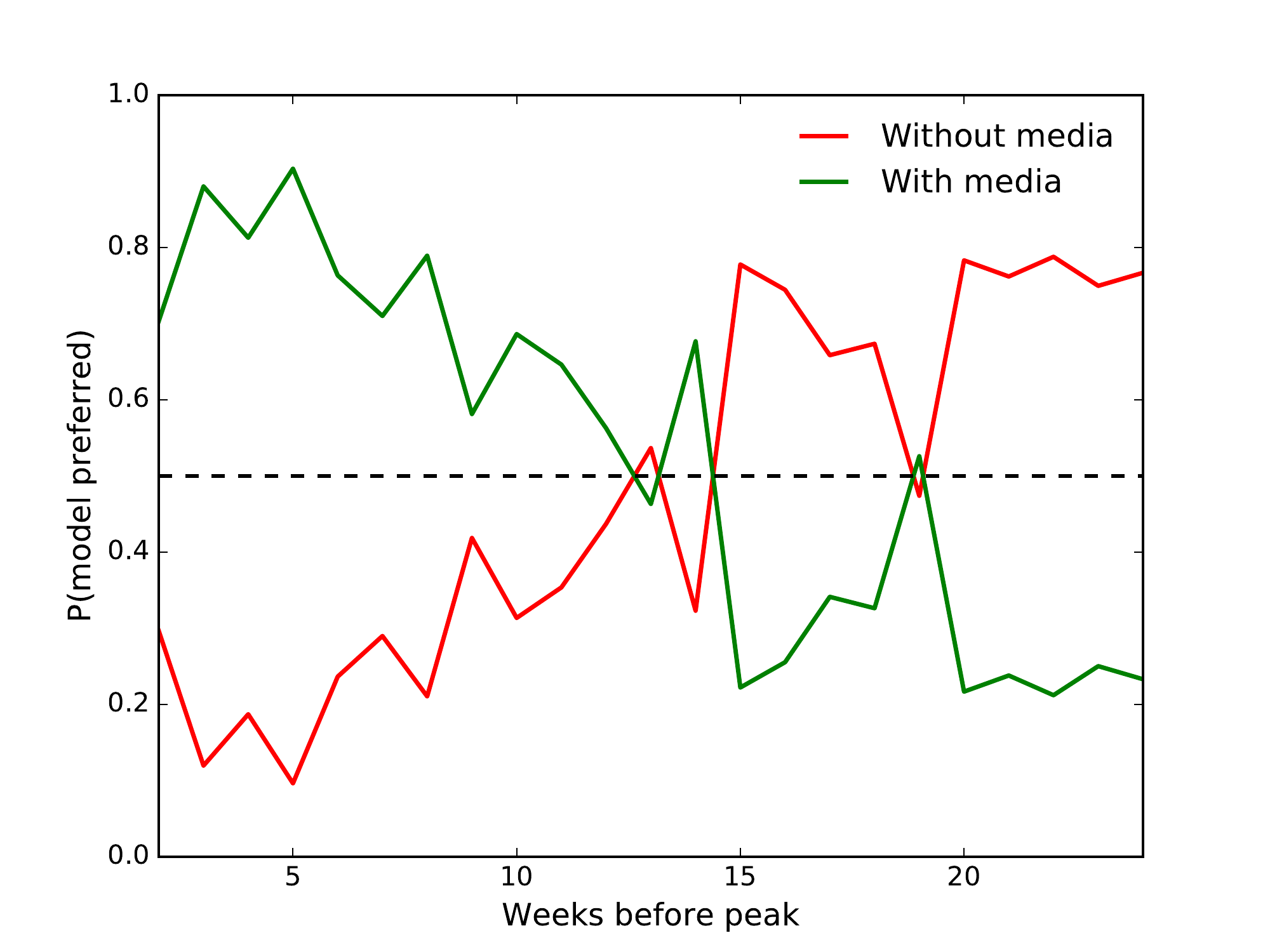}
  \caption{Quality of model fits as a function of lead time.
      Probability that the no-media model (green) and media model with $f_m(I) = 1 - p_m I$ (blue) are the better model as a function of number of weeks before peak.
      In each case we fit models to data over a 16-week period.
      The 2009/10 pandemic influenza season has been excluded.}
      \label{fig:pBothLeadTime}
      \end{figure}

We remark that much of the improvement made by the media function $f_m$ comes from better describing the post-peak period.
The no media (i.e., $f(I)\equiv 1$) model becomes preferable as more of the data leading up to the peak of each season is used to fit the models.
In Figure \ref{fig:pBothLeadTime} we show the average conditional probability of selecting each model
as a function of the number of weeks of data used before the peak.
The no media and $f_m$ models are always preferred over the other media function models (i.e., using $f_1$, $f_2$ and $f_3$),
with the $f_m$ model being preferred up until around 10-12 weeks before the peak.
When fitting data earlier than 12 weeks before the peak the no media model is preferred,
suggesting that the effect of media coverage becomes more important later in the season.
Furthermore,
neither model is able to reliably predict the peak of the infection in terms of either size or timing 
based upon data from before the peak only.
This suggests that in order to make accurate predictions and estimate parameters rather than explain an existing data set when only small amounts of data are available,
we must use a more advanced methodology such as data assimilation \cite{Shaman2012}.

% \note{both models are not so great at predicting the peak using only data from before the peak. perhaps an example figure showing this.}

\section{Discussion}
\label{sec:discussion}

Mass media is clearly an important tool for changing peoples' behaviour during disease outbreaks.
A better understanding of the relationship between media coverage of outbreaks and subsequent behavioural change can aid mathematical modelling efforts,
as well as the development of public policy around the best use of this resource to inform the public and control the spread of a disease.

By using data collected from Twitter,
we have proposed a new, simple media function
to describe the reduction in disease transmission due to media effects.
When incorporated into a deterministic SEEIIR model,
this media function describes incidence data better than a model without media effects, and better than previously proposed media functions.
%\note{moar on specifics of predicting peak time and size}
We observed a relationship between outbreak size and media awareness,
with a quadratic model becoming more likely as the final size of the outbreak increased.
This suggests that the relationship between media coverage and infection rates is nonlinear,
especially in more severe seasons.
Future extensions to the media function could incorporate extra reductions in disease transmission due to factors such as early media coverage,
pre-existing immunity,
or seasonality.
Public awareness campaigns could lead to an increase in early-season social media activity and sharing of news articles,
and could be implemented in the current model via a time lag.
Indeed,
we observed such an effect for the 2014/15 season where changes in retweeting activity preceded ILI rates by a number of weeks.
Mass media campaigns have been shown to increase flu-related hospital visits \cite{Codish2014} and vaccination rates \cite{Ma2006,Yoo2010}.
It is further possible that any potential reduction in transmission in one season due to the effects of mass media could decrease pre-existing immunity for the next season,
an effect which could be modelled by conditioning the media function on the total amount of media engagement from the previous season.
Identifying any such potential process is of course confounded by the presence of multiple influenza strains circulating in any particular season with differing levels of pre-exisitng immunity;
modelling such a hierarchy of time-lagged effects requires a more sophisticated strategy and is left for future work.

The interplay between mass media,
social network influence,
human behavioural change,
and disease transmission is complex,
and this work merely scratches the surface of the processes which could be modelled using this framework.
Further extensions could build upon efforts to incorporate interactions between social and contact network structures into the model \cite{Funk2012}
by inferring the mass media effect directly from social network data.
There is also an emerging body of work around using open data to infer human behaviours such as mobility patterns \cite{Frank2013} and voluntary avoidance \cite{Bayham2015}.
The same data used here to track media engagement could potentially be exploited to quantify such effects,
as well as to develop a proxy for real-time surveillance on practices such as vaccination,
which we aim to incorporate into future refinements of this model.

A critical assumption made in this work is that the population is homogeneously mixing and not age-stratified.
This is of course far from being the case for Twitter users -- 
indeed, it is well-known that the demographics of Twitter use in the United States are biased towards 
adults aged 18-29, 
African-Americans, 
and
urban residents \cite{Duggan2013},
and word usage has been shown to correlate with a number of socioeconomic and health characteristics \cite{Mitchell2013,Alajajian2015}.
Despite these biases,
the roughly 10\% of American adults who are estimated to use Twitter represents a far larger sample size than those of traditional surveys.
Furthermore, 
for simplicity 
and because the keywords we used were sufficiently specific,
we did not filter tweets for relevance.
Manual examination of a sample of tweets indicated that an insignificant number of tweets were misclassified as being about influenza,
however constraining the tweet corpus may lead to further improvements in the results.

This work fits into a growing field of research on disease prediction using open data \cite{Althouse2015},
particularly from social network usage.
Great advances have already been made on algorithms to predict rare and seasonal diseases, 
especially in the computer science literature \cite{Chen2014}.
Our results represent a first attempt at incorporating this emerging data stream into more traditional modelling efforts,
and hopefully at better understanding the interactions between media and disease dynamics.

\clearpage

\section*{Acknowledgment}

The authors wish to thank PS Dodds and CM Danforth from the Computational Story Lab at the University of Vermont for use of the Twitter Gardenhose feed for this study.

\section*{Research ethics}

Does not apply.

\section*{Animal ethics}

Does not apply.

\section*{Permission to carry out fieldwork}

Does not apply.

\section*{Data Availability}

Data will be available via Dryad: \url{http://dx.doi.org/10.5061/dryad.593cc}

\section*{Competing Interests}

We have no competing interests.

\section*{Authors' Contributions}

LM and JVR conceived of the study; LM performed data analysis and simulations; LM and JVR wrote the manuscript.

\section*{Funding}

The authors received funding from Data To Decisions Cooperative Research Centre (D2D CRC).
JVR received funding from the Australian Research Council through the Centre of Excellence for Mathematical and Statistical Frontiers (ACEMS), 
the Future Fellowship scheme (FT130100254), 
and the National Health and Medical Research Council (NHMRC)
Centre of Research Excellence for 
Policy Relevant Infectious Disease Simulation and Mathematical Modelling (PRISM$^2$).

%%%%%%%%%% Insert bibliography here %%%%%%%%%%%%%%
\bibliographystyle{vancouver}
\bibliography{bmc_article}      % Bibliography file (usually '*.bib' )

\clearpage

\section*{Appendix}
\appendix

\begin{figure}[h!]
\includegraphics[width=\columnwidth]{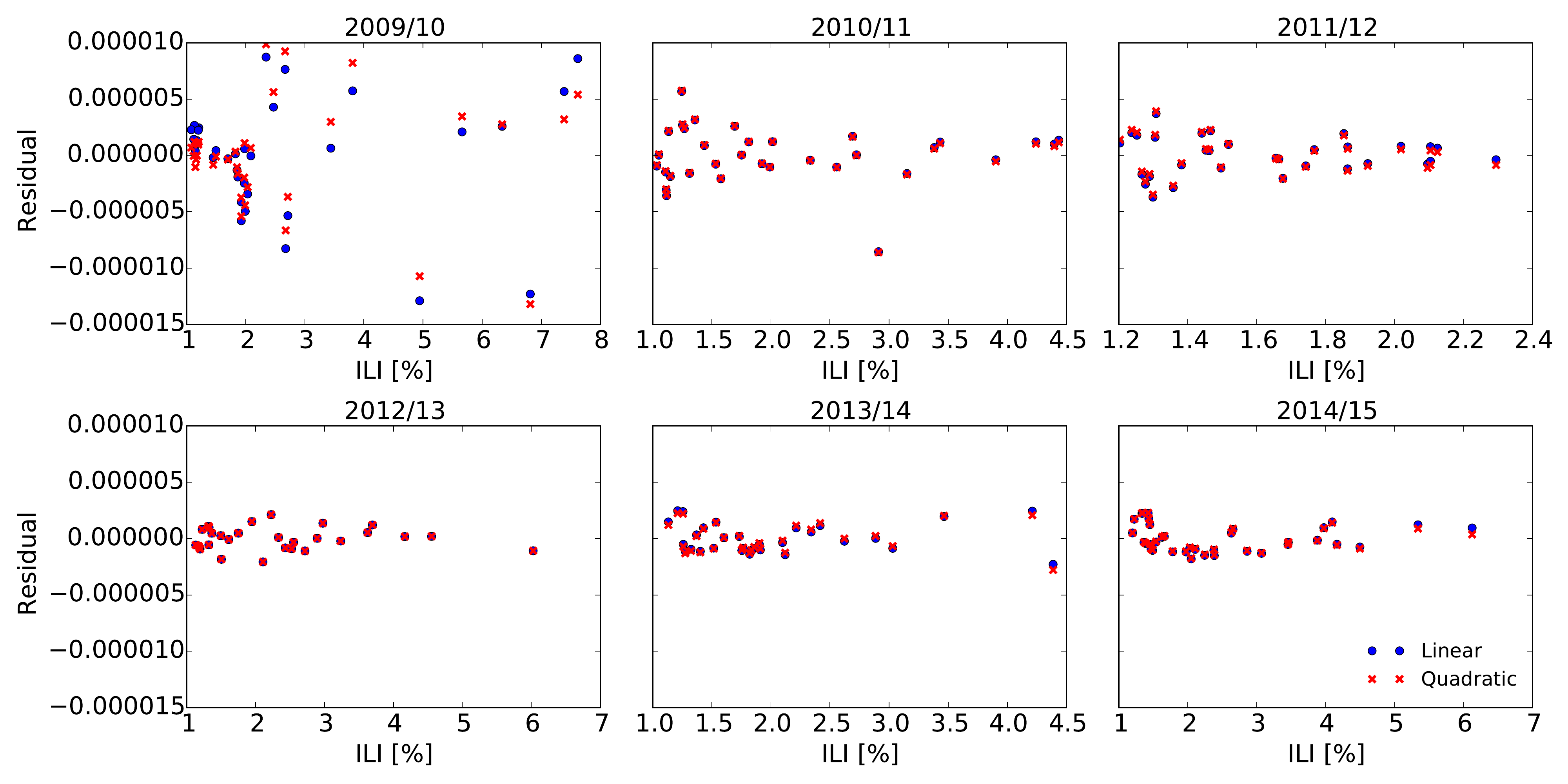}
\caption{Residual plots for linear (blue dots) and quadratic (red crosses) media function models.}
\label{fig:residualPlots}
\end{figure}

\begin{figure}[h!]
\includegraphics[width=0.7\columnwidth]{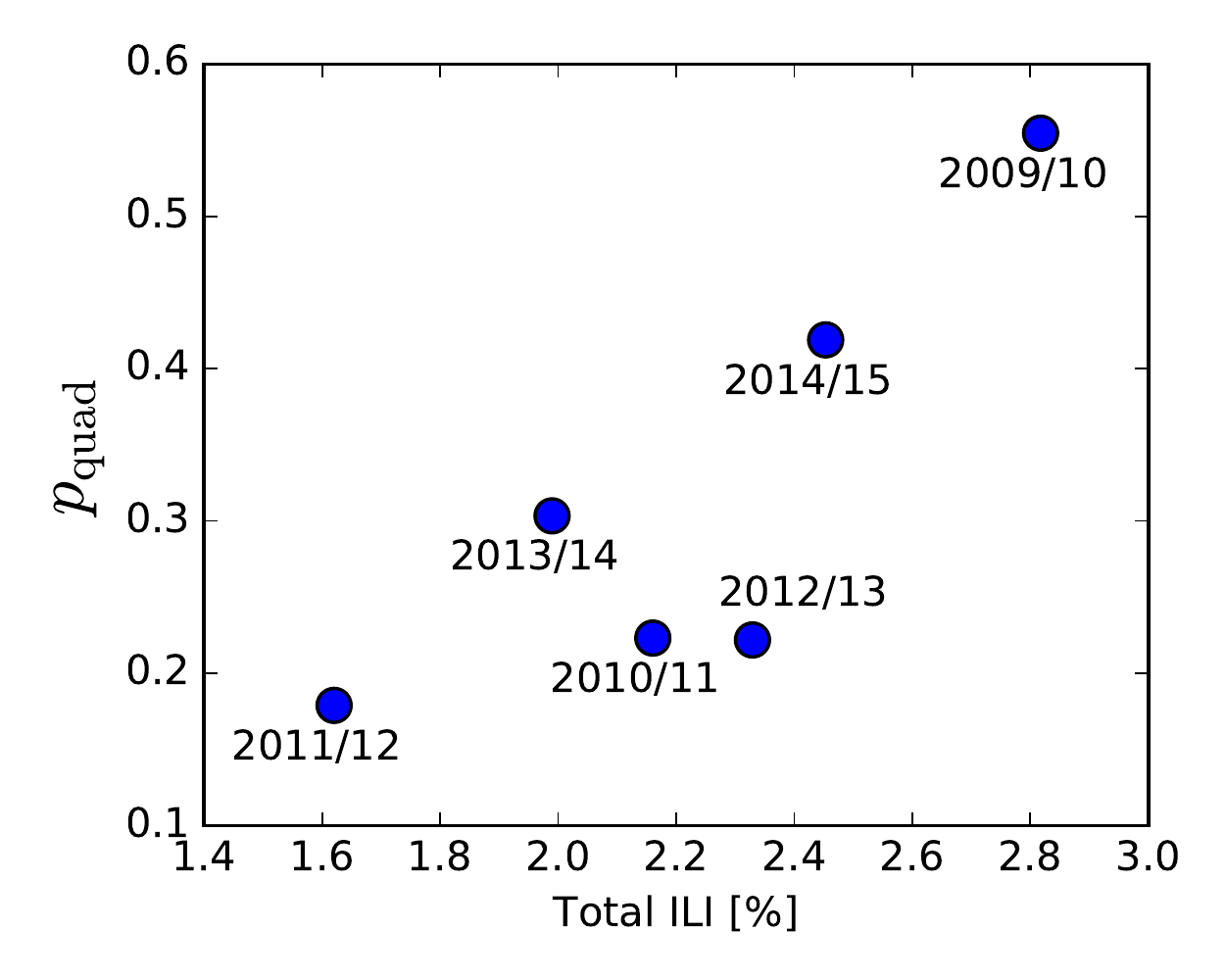}
\caption{Relationship between relative Akaike weight for the quadratic model $p_{\rm quad}$ and the observed total ILI for each influenza season 2009/10-2014/15.}
\label{fig:pQuad}
\end{figure}

In Figure \ref{fig:residualPlots} we show the residuals from linear (blue circles)
and quadratic (red crosses) least-square fits to the retweet-infected data from Figure~1 of the main text,
for all seasons 2009/10-2014/15 considered in this study.
Table \ref{tbl:linquad} shows the relative Akaike weights for the linear and quadratic models for each year,
where the relative weights $p_i$ are given by
\[
p_i =  \exp\left(\frac{AIC_{\rm min} - AIC_i}{2}\right)
\]
and $AIC_i$ are the AIC values for each model, $AIC_{\rm min}$ is the minimum value of $AIC_i$.
These weights represent the probability that the $i$th model minimises the information loss in describing the data,
relative to the best-fitting model \cite{Anderson2008}.
Figure \ref{fig:pQuad} shows relative Akaike weight of the quadratic model as a function of final outbreak size, 
suggesting that nonlinear effects will increasingly come into play for more severe outbreaks and pandemics.

\begin{table}[h!]
\caption{Relative Akaike weights $p_{\rm lin}$ and $p_{\rm quad}$ for selecting the linear and quadratic models during each influenza season 1998/99-2014/15. Weights for each preferred model are shown in bold.}
\label{tbl:linquad}
      \begin{tabular}{lcc}
        \hline
          year &  $p_{\rm lin}$ & $p_{\rm quad}$  \\ 
           \hline
        2009/10 & 0.4454 & {\bf 0.5546} \\
        2010/11 & {\bf 0.7769} & 0.2231 \\
        2011/12 & {\bf 0.8211} & 0.1789 \\
        2012/13 & {\bf 0.7781} & 0.2219 \\
        2013/14 & {\bf 0.6966} & 0.3034 \\
        2014/15 & {\bf 0.5811} & 0.4189 \\
        \hline
      \end{tabular}
\end{table}

\begin{figure}[h!]
  \includegraphics[width=\columnwidth]{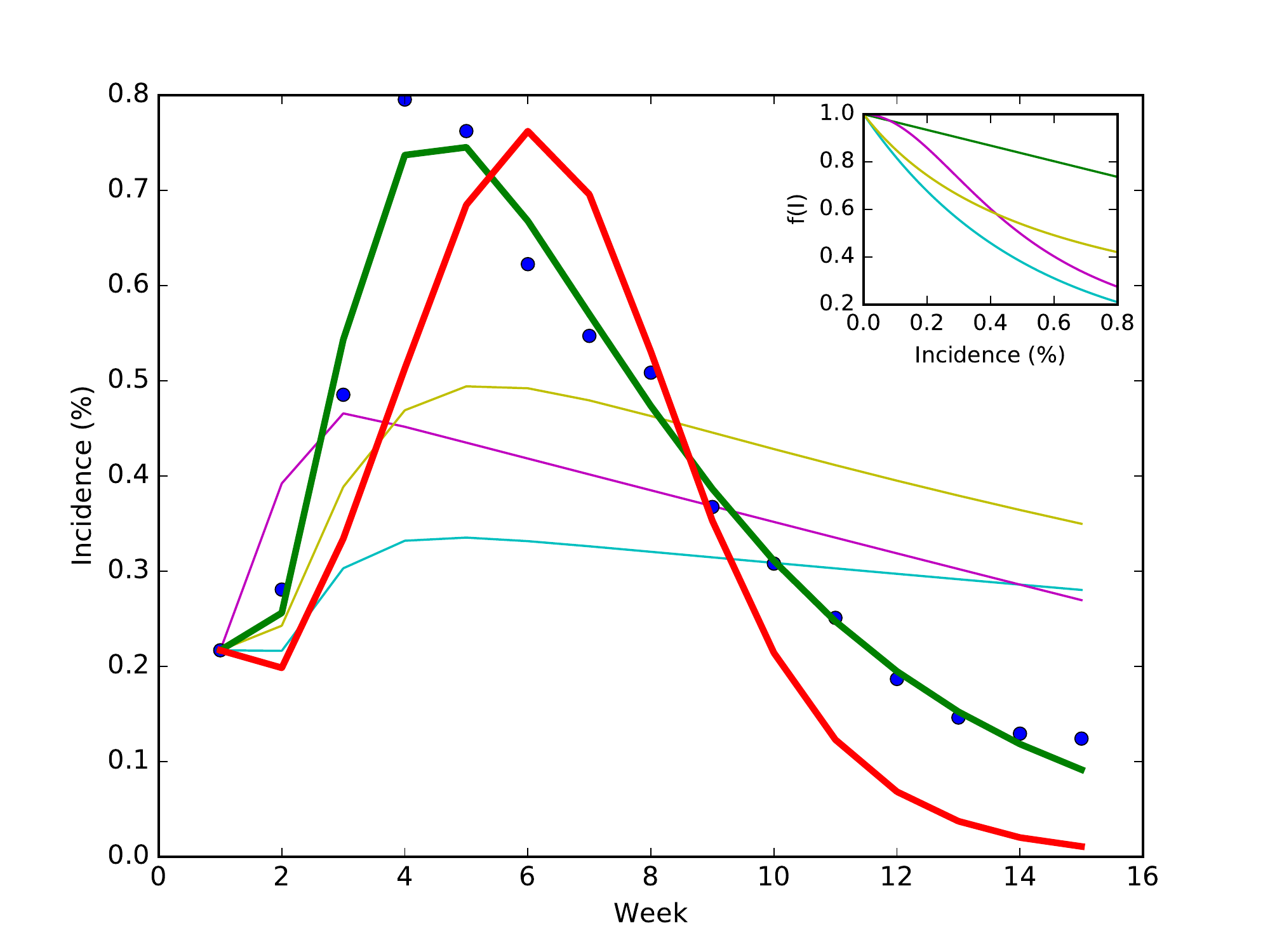}
  \caption{Best model fits to data from 2013/14 flu season.
      Weekly influenza data (blue dots), and model fits for SEIR model without media function (red), with media function (green), and variations 
$f_1$ (cyan),
$f_2$ (magenta)
and $f_3$ (yellow),
for the 2013/14 influenza season (USA). (cf. Figure 3 from the main text.)}
      \label{fig:2013_SEIR}
      \end{figure}

Table \ref{tbl:allParams} shows the best-fitting parameters for the SEEIIR and SEEIIR-M models for all seasons 1998/99-2014/15,
analogously to Table 1 in the main text.
For completeness we also present results where we have used a standard SEIR model with only one exposed and infectious compartment each.
Table \ref{tbl:probs_SEIR} is an analog of Table 2 from the main text, showing the average conditional probability of selecting each model,
and Figure \ref{fig:2013_SEIR} shows example fits to observations of the percentage of new laboratory-confirmed influenza cases per week for the SEIR and SEEIIR-M models,
similarly to Figure 3 in the main text.
In terms of model fitting,
 the confidence intervals in Table \ref{tbl:probs_SEIR} show no significant difference in the likelihood of selecting each media model.
Figure \ref{fig:2013_SEIR} suggests that in 2013/14 the media functions $f_1$, $f_2$ and $f_3$ fit the data slightly worse using an SEIR model than with an SEEIIR model,
while the $f(I)\equiv 1$ and $f_m$ fit describe the data essentially as well using an SEIR model than with an SEEIIR model.

%%%%% SEEIIR model %%%%%        
\begin{longtable}{l|l|lllll}
        \caption{Parameters of best fit for SEEIIR and SEEIIR-M models for all influenza seasons.}
	\label{tbl:allParams}\\

        \hline
        	& model	& $R_0$   & $1/\sigma$	& $1/\gamma$	& $p_i$	& $p_{AIC}$ \\ 
	& 		& 		& (days)		& (days)		&		& \\
	\hline
	\endfirsthead

        \hline
        	& model	& $R_0$   & $1/\sigma$	& $1/\gamma$	& $p_i$	& $p_{AIC}$ \\ 
	& 		& 		& (days)		& (days)		&		& \\
	\hline
	\endhead

	\hline \multicolumn{7}{r}{{Continued on next page}} \\ \hline
	\endfoot

	\hline \hline
	\endlastfoot

\hline
& $f(I) \equiv 1$ & 1.17 & 1.00  & 2.41 & -- & 0.80 \\
& $f_m$ & 2.00 & 1.92  & 1.41 & 0.22 & 0.02 \\
1998 & $f_1$ & 1.30 & 1.00  & 2.30 & 0.02 & 0.15 \\
& $f_2$ & 1.35 & 1.21  & 2.06 & 0.06 & 0.03 \\
& $f_3$ & 1.58 & 1.98  & 2.02 & 5.71 & 0.00 \\
\hline
& $f(I) \equiv 1$ & 1.23 & 1.00  & 1.63 & -- & 0.00 \\
& $f_m$ & 1.47 & 1.28  & 1.58 & 0.11 & 0.99 \\
1999 & $f_1$ & 1.63 & 1.71  & 1.38 & 0.03 & 0.01 \\
& $f_2$ & 2.00 & 1.95  & 1.92 & 0.62 & 0.00 \\
& $f_3$ & 1.76 & 1.90  & 1.30 & 4.14 & 0.01 \\
\hline
& $f(I) \equiv 1$ & 1.14 & 1.00  & 1.26 & -- & 0.00 \\
& $f_m$ & 1.30 & 1.00  & 1.23 & 0.21 & 1.00 \\
2000 & $f_1$ & 1.35 & 1.13  & 1.13 & 0.04 & 0.00 \\
& $f_2$ & 1.33 & 1.99  & 1.97 & 1.05 & 0.00 \\
& $f_3$ & 1.59 & 1.99  & 1.99 & 1.06 & 0.00 \\
\hline
& $f(I) \equiv 1$ & 1.10 & 1.00  & 1.98 & -- & 0.00 \\
& $f_m$ & 1.45 & 1.14  & 1.33 & 0.24 & 0.89 \\
2001 & $f_1$ & 1.34 & 1.37  & 1.32 & 0.03 & 0.00 \\
& $f_2$ & 1.38 & 1.29  & 1.37 & 0.26 & 0.00 \\
& $f_3$ & 1.46 & 1.07  & 1.38 & 3.15 & 0.11 \\
\hline
& $f(I) \equiv 1$ & 1.13 & 1.00  & 1.00 & -- & 0.00 \\
& $f_m$ & 1.28 & 1.00  & 1.00 & 0.29 & 1.00 \\
2002 & $f_1$ & 1.56 & 1.61  & 1.55 & 0.19 & 0.00 \\
& $f_2$ & 1.28 & 1.35  & 1.00 & 0.70 & 0.00 \\
& $f_3$ & 1.71 & 1.83  & 1.74 & 0.54 & 0.00 \\
\hline
& $f(I) \equiv 1$ & 1.17 & 1.00  & 1.46 & -- & 1.00 \\
& $f_m$ & 1.17 & 1.00  & 1.46 & 0.00 & 0.00 \\
2003 & $f_1$ & 1.17 & 1.00  & 1.46 & 0.00 & 0.00 \\
& $f_2$ & 1.17 & 1.00  & 1.46 & 0.00 & 0.00 \\
& $f_3$ & 1.19 & 1.00  & 1.46 & 56.78 & 0.00 \\
\hline
& $f(I) \equiv 1$ & 1.11 & 1.00  & 1.30 & -- & 0.00 \\
& $f_m$ & 1.36 & 1.02  & 1.00 & 0.30 & 1.00 \\
2004 & $f_1$ & 2.00 & 2.00  & 2.07 & 0.68 & 0.00 \\
& $f_2$ & 1.23 & 1.16  & 1.10 & 0.41 & 0.00 \\
& $f_3$ & 1.14 & 1.09  & 1.20 & 33.94 & 0.00 \\
\hline
& $f(I) \equiv 1$ & 1.10 & 1.00  & 1.09 & -- & 0.00 \\
& $f_m$ & 1.24 & 1.00  & 1.01 & 0.28 & 1.00 \\
2005 & $f_1$ & 1.10 & 1.78  & 2.95 & 0.14 & 0.00 \\
& $f_2$ & 1.33 & 1.00  & 1.00 & 2.58 & 0.00 \\
& $f_3$ & 1.19 & 1.00  & 1.29 & 4.23 & 0.00 \\
\hline
& $f(I) \equiv 1$ & 1.12 & 1.00  & 1.22 & -- & 0.00 \\
& $f_m$ & 1.40 & 1.00  & 1.00 & 0.36 & 1.00 \\
2006 & $f_1$ & 1.32 & 1.04  & 1.16 & 0.05 & 0.00 \\
& $f_2$ & 1.62 & 1.04  & 1.00 & 2.51 & 0.00 \\
& $f_3$ & 1.54 & 1.78  & 1.00 & 1.24 & 0.00 \\
\hline
& $f(I) \equiv 1$ & 1.12 & 1.00  & 1.64 & -- & 0.01 \\
& $f_m$ & 1.26 & 1.00  & 1.42 & 0.13 & 0.99 \\
2007 & $f_1$ & 1.45 & 1.49  & 1.09 & 0.04 & 0.00 \\
& $f_2$ & 1.72 & 1.27  & 1.00 & 0.66 & 0.00 \\
& $f_3$ & 1.42 & 1.84  & 1.84 & 2.11 & 0.00 \\
\hline
& $f(I) \equiv 1$ & 1.14 & 2.83  & 1.06 & -- & 0.00 \\
& $f_m$ & 1.28 & 1.00  & 1.07 & 0.27 & 0.96 \\
2008 & $f_1$ & 1.40 & 1.13  & 1.00 & 0.06 & 0.00 \\
& $f_2$ & 1.25 & 1.02  & 1.02 & 0.57 & 0.04 \\
& $f_3$ & 2.00 & 1.18  & 1.13 & 0.48 & 0.00 \\
\hline
& $f(I) \equiv 1$ & 1.16 & 1.00  & 1.74 & -- & 0.00 \\
& $f_m$ & 1.10 & 1.10  & 3.39 & 0.06 & 0.00 \\
2009 & $f_1$ & 1.39 & 1.28  & 1.46 & 0.03 & 0.91 \\
& $f_2$ & 1.41 & 1.64  & 1.19 & 0.11 & 0.00 \\
& $f_3$ & 1.53 & 1.57  & 1.28 & 4.83 & 0.09 \\
\hline
& $f(I) \equiv 1$ & 1.11 & 1.00  & 1.31 & -- & 0.35 \\
& $f_m$ & 1.16 & 1.28  & 1.04 & 0.06 & 0.00 \\
2010 & $f_1$ & 1.30 & 1.86  & 2.32 & 0.22 & 0.00 \\
& $f_2$ & 1.28 & 1.02  & 1.01 & 0.33 & 0.56 \\
& $f_3$ & 1.21 & 1.00  & 1.16 & 7.44 & 0.09 \\
\hline
& $f(I) \equiv 1$ & 1.10 & 2.75  & 1.09 & -- & 0.00 \\
& $f_m$ & 1.24 & 1.00  & 1.00 & 0.48 & 1.00 \\
2011 & $f_1$ & 1.57 & 1.00  & 1.33 & 0.38 & 0.00 \\
& $f_2$ & 1.10 & 1.03  & 1.42 & 0.35 & 0.00 \\
& $f_3$ & 1.25 & 1.09  & 1.56 & 1.24 & 0.00 \\
\hline
& $f(I) \equiv 1$ & 1.15 & 1.00  & 1.94 & -- & 0.00 \\
& $f_m$ & 1.44 & 1.30  & 1.70 & 0.21 & 0.96 \\
2012 & $f_1$ & 1.57 & 1.63  & 1.51 & 0.07 & 0.01 \\
& $f_2$ & 1.41 & 1.51  & 1.57 & 0.22 & 0.00 \\
& $f_3$ & 1.54 & 1.44  & 1.61 & 2.96 & 0.03 \\
\hline
& $f(I) \equiv 1$ & 1.16 & 1.00  & 1.25 & -- & 0.00 \\
& $f_m$ & 1.53 & 1.78  & 1.00 & 0.34 & 1.00 \\
2013 & $f_1$ & 1.68 & 2.13  & 1.01 & 0.07 & 0.00 \\
& $f_2$ & 1.33 & 1.33  & 1.19 & 0.35 & 0.00 \\
& $f_3$ & 1.75 & 1.23  & 1.13 & 1.19 & 0.00 \\
\hline
& $f(I) \equiv 1$ & 1.19 & 1.00  & 2.46 & -- & 0.00 \\
& $f_m$ & 1.70 & 1.67  & 2.00 & 0.18 & 0.81 \\
2014 & $f_1$ & 1.65 & 2.00  & 2.02 & 0.06 & 0.00 \\
& $f_2$ & 1.57 & 1.68  & 2.02 & 0.12 & 0.02 \\
& $f_3$ & 1.96 & 1.93  & 1.85 & 2.64 & 0.17 \\

\hline
\end{longtable}

\begin{table}[h!]
\caption{Average probability of selecting each SEIR-type model over the 1998/99-2014/15 seasons.
(cf. Table 2 of the main text.)
The 2009/10 pandemic influenza season has been excluded.}
\label{tbl:probs_SEIR}
      \begin{tabular}{lcc}
        \hline
           &  $p_{AIC}$ & 95\% CI  \\ 
           \hline
        $f(I)\equiv 1$                       & 0.0635 & [0,0.1714] \\
        $f_m(I) = 1 - p_m I$            & 0.9122 & [0.7974,1] \\
        $f_1(I) = \exp(-\gamma p_1 I)$   & 0.0010 & [0,0.0026] \\
        $f_2(I) = \frac{1}{1 + p_2 I^2}$ & 0.0134 & [0,0.0388] \\ 
        $f_3(I) = \frac{1}{1 + p_3 I}$   & 0.0100 & [0,0.0255]\\
        \hline
      \end{tabular}
\end{table}

\end{document}